\documentclass[a4paper,fleqn,usenatbib,useAMS]{mn2e}
\usepackage{graphicx}	
\usepackage[flushleft]{threeparttable}
\usepackage{amsmath}	
\usepackage{amssymb}	
\usepackage{multicol}        
\usepackage{bm}		
\usepackage{pdflscape}	
\usepackage{color}
\usepackage{hyperref}
\usepackage{natbib}
\usepackage[T1]{fontenc}
\usepackage{ae,aecompl}
\usepackage{newtxtext,newtxmath}
 
\def\aap{AA}

\def\mnras{MNRAS}
\def\apj{ApJ}

\def\aj{AJ}

\def\procspie{Proc.~SPIE}
\def\nat{Nat}

\title[Quasar lenses in DES-DR1]{Quasar Lenses in the South: searches over the DES public footprint}

\author[ A.~Agnello \& C.~Spiniello]{Adriano Agnello$^{1}$, Chiara Spiniello$^{1,2}$ 
\\
$^{1}$European Southern Observatory, Karl-Schwarzschild-Strasse 2, 85748 Garching, Germany \\
$^{2}$INAF - Osservatorio Astronomico di Capodimonte, Salita Moiariello, 16, I-80131 Napoli, Italy \\
}
\date{Last updated 2018 May 24}
\pubyear{2018}
\hypersetup{draft}

\begin{document}
\label{firstpage}
\pagerange{\pageref{firstpage}--\pageref{lastpage}}
\maketitle

\begin{abstract}
We have scanned 5000 deg$^{2}$ of Southern Sky to search for strongly lensed quasars
with five methods, all source-oriented, but based on different assumptions and selection criteria. We analyse morphological searches based on Gaia multiplet detection and chromatic offsets, fibre-spectroscopic preselection, and X-ray and radio preselection.
The performance and complementarity of the methods are evaluated on a common sample of known lenses in the Dark Energy Survey public DR1 footprint. 
We recovered in total 13 known lenses, of which 8 quadruplets. The method that found the largest number of known lenses is the one based on morphological and colour selection of objects from the WISE and Gaia-DR2 Surveys. 
We finally present a list of high-grade candidates from each method, to facilitate follow-up spectroscopic campaigns, including two previously unknown quadruplets: WG210014.9-445206.4 and WG021416.37-210535.3.

\end{abstract}


\begingroup
\let\clearpage\relax
\endgroup
\newpage

\section{Introduction}
\label{intro}
Sizeable samples of gravitationally lensed quasars, spanning a range in redshift, image-separation and mass, are essential to various scientific purposes. 
However these objects are extremely rare: according to \citet[hereafter OM10]{Oguri10}, one over $\approx10^4$ quasars is strongly lensed.
%
One of the first efforts to build a large and well-studied statistical sample of strong gravitational lensed quasars has been made by The Cosmic Lens All-Sky Survey (CLASS, \citealt{Myers03}). CLASS has obtained high-resolution radio images from the Very Large Array (VLA, at 8.4 GHz) of over 13000 flat-spectrum radio sources, finding $22$ radio-loud lenses.  However, in radio lens surveys like CLASS, the redshift distribution of the flat-spectrum  sources is poorly constrained (e.g., \citealt{Munoz03}).
%
Only with the advent of the Sloan Digital Sky Survey (SDSS, \citealt{York00}, which in its 14th Data Release identified 526,356 quasars (\citealt{Paris17}), has it been possible to construct a large sample of lensed quasars also in the optical (e.g., 
The Sloan Digital Sky Survey Quasar Lens Search, SQLS, \citealt{Oguri06}), beyond the bright lenses identified in older searches.

More recently, thanks to new imaging sky surveys, with wide footprints and sufficient depth and image quality, like ATLAS \citep{Shanks15}, DES \citep{DES05,DES16} and KiDS \citep{deJong15,deJong17}, the search for lensed quasars has received new impetus. 
%
Nevertheless, to this day, a big portion of the Southern Sky still remains unexplored and the spatial density on sky of known lenses remains much higher at dec.$>0$. 

In order to fully exploit the wealth of data from different surveys, a suite of new methods to search for lensed quasars (QSOs) have been developed, based on: morphology and visual inspection (\citealt{Lin17, Diehl17}); spectroscopy (e.g. \citealt{Oguri06,Inada12,More16}); data mining on catalog magnitudes (\citealt{Agnello15a,Williams17,Agnello17met}); and variability \citep{Berghea17}.  

As recently quantified by Spiniello et al. (2018, subm.), different search methods are somewhat complementary, as each search relies on different criteria and ancillary information. The application of multiple, state-of-the-art methods over the same footprint then serves two scopes: assessing the role of selection bias in lens samples; and maximizing purity and completeness in lens searches.
%

With this paper, we perform a comparison study over the footprint covered by the Dark Energy Survey DR1 (DES, \citealt{DES16, Abbott18}). Our choice is motivated by three criteria: the DES-DR1 release is public, which ensures reproducibility; its footprint lies mostly in the South, which is still largely unexplored; and the sheer convenience of display from the NCSA DESaccess cutout service, which enabled a swift visual inspection of image cutouts.
Magnitudes $W1,W2,W3$ from the Wide-field Infrared Survey Explorer (hereafter WISE; \citealt{Wright10}) and from the Gaia space mission (\citealt{Gaia16}) are given in their native Vega system.

\begin{figure}
\includegraphics[width=0.45\textwidth]{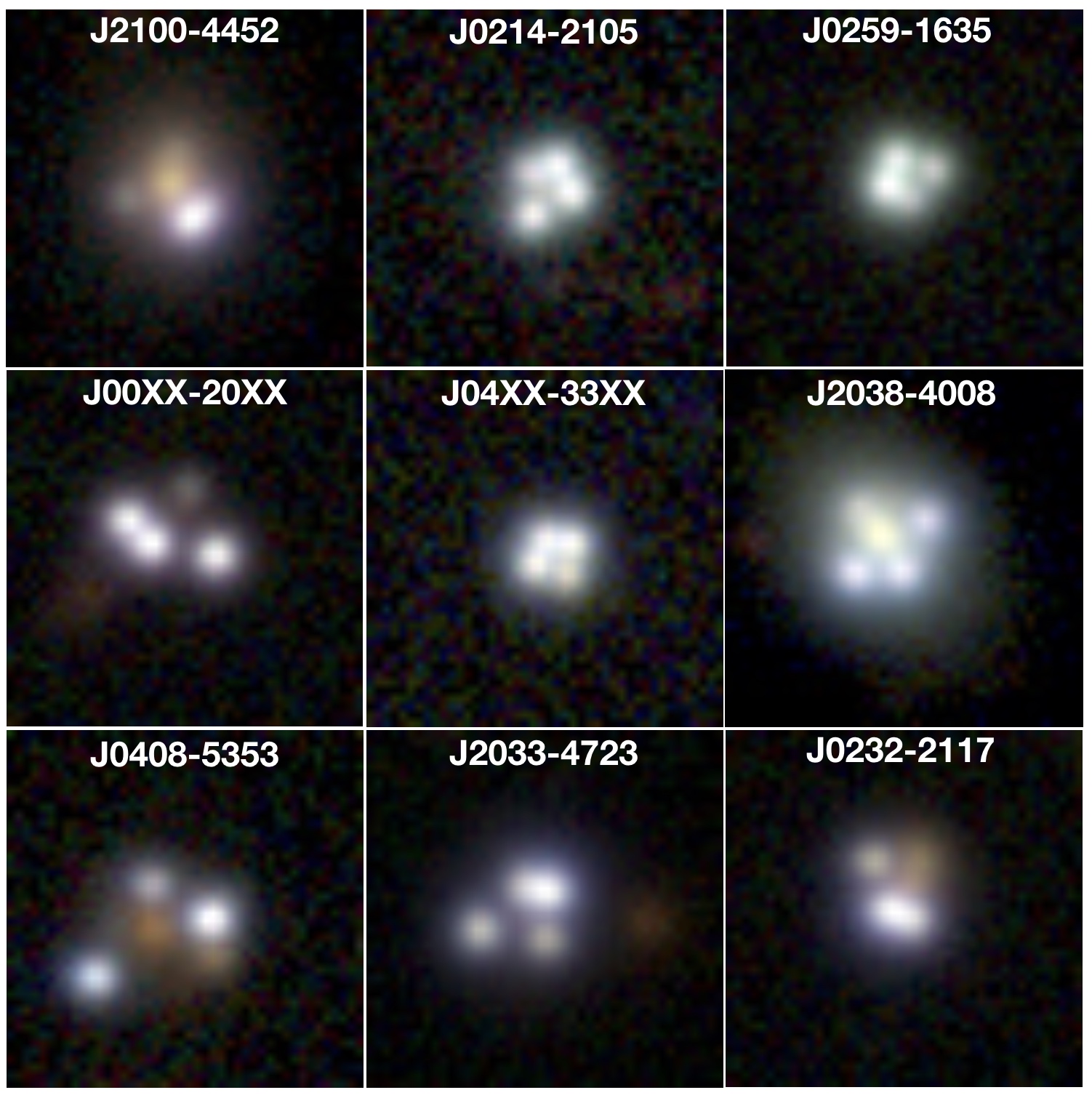}
 \caption{The nine quadruplets found by at least one method in the DES public footprint, with WISE extragalactic colours. The 11.5$" \times$11.5$"$ cutouts are obtained  from the DES Data Access page. WG~2100-4452 is a previously unknown quad, and the discovery of WG~0214-2105 has been recently reported by Agnello (2018, RNAAS). WG~00-20 and WG~04-33 have been found independently in STRIDES and are awaiting publication (Schechter \& Treu, private comm.). Table~1 gives references for all quadruplets.}
\label{fig:quads}
\end{figure}

The paper is organized as follows.
In Section~\ref{methods} we describe each of the search methods, and its possible biases. 
In Section~\ref{performance} we estimate the performance of each method, in terms of known lens recovery and of new candidates produced over the DES-DR1 footprint. We also provide tables of candidates, to facilitate spectroscopic follow-up, including at least two `new' quadruplets: WG~210014.9-445206.4 and WG~021416.37-210535.3 (figure \ref{fig:quads}). 
We conclude in Section~\ref{results}.

%
%
\section{The Methods}
\label{methods}
%

We use five different methods, of which: two are mostly morphological and rely on Gaia multiplet detection or chromatic offsets; one is based on spectroscopic preselection from fibre-spectroscopic surveys; and two are based on radio and X-ray preselection. 

The searches are structured similarly: a first step of \textit{object preselection}, based on shallow cuts on catalog magnitudes and colours; a second step of \textit{target} selection, still based on catalog entries; and a final step of \textit{candidate} selection, through visual inspection of image cutouts. 
For the sake of convenience, the DES-DR1 public cutout service was used. However, in its current version, it properly displays between 50\% and 70\% of the cutouts that are produced. For this reason, the final numbers of candidates given in this paper should be taken as purely indicative, and a significant number of quasar lenses may still be found among the systems that were not properly displayed.



\subsection{BaROQuES: Blue and Red Offsets of Quasars and Extragalactic Sources}
\label{baroques}
The Gaia mission had $\approx2^{\prime\prime}$ image resolution in DR1 and $\approx0.5^{\prime\prime}$ in DR2, which enables the recognition of multiple images in lensed quasars that are otherwise not deblended by the image-processing pipelines of ground-based surveys.
The search of quasar lenses in DES through Gaia-DR1 multiplet recognition has already been made within STRIDES \citep{Agnello17}. However, not all known lenses are deblended into multiple \texttt{source} entries in Gaia-DR1 (e.g. \citealt{Agnello17}).

For systems that are not Gaia multiplets, we can still use the accurate astrometry of Gaia to compare chromatic shifts in image centroids among different surveys, under the hypothesis that lenses have non-negligible offsets due to the different contributions of source and deflector in different bands. To this aim, we consider objects that are bright enough to have positions in the 2MASS catalog, and consider their 2MASS-Gaia offsets. Matters are slightly complicated, due to atmospheric differential refraction (ADR) and the choice of astrometric calibrators for each survey. Recent searches based on chromatic offsets \citep{Lemon18} relied on SDSS-vs-Gaia solutions, calibrated on astrometric reference stars, and had abundant contamination from isolated (unlensed) quasars. This is due to the different colours of quasars and astrometric stars, which can result in $\approx0.3^{\prime\prime}$ offsets due to ADR (with typical $\approx20$~deg Zenith angles of ground-based surveys).

We resolve this by means of field-corrected offsets, using our own BaROQuES scripts\footnote{Blue and Red Offsets of Quasars and Extragalactic Sources, available upon request, \texttt{https://github.com/aagnello} .}, as follows. 
First, we work with a sample of quasar-like objects, selected in WISE as:
\begin{eqnarray}
W1-W2>0.35-\sqrt{(\delta W1)^{2}+(\delta W2)^{2}}\ ,\\
W1-W2>0.2+\sqrt{(\delta W1)^{2}+(\delta W2)^{2}}\ ,\\
W2-W3>2.1+\sqrt{(\delta W2)^{2}+(\delta W3)^{2}}\ ,\\
W1<17\ ,\ W2<15.6\ ,\ W3<11.8\ .
\end{eqnarray}
Then, for each given object we examine a surrounding patch of $1.0\times1.0\mathrm{deg}^{2},$ containing $\approx40$ QSO-like neighbours. We then compute the average offsets between their 2MASS and Gaia coordinates $\delta x=-\cos(\mathrm{dec.})\delta\mathrm{r.a.},$
 $\delta y=\delta\mathrm{dec.},$ 
subtract them from the offsets of the given object, and retain it only if this corrected offset is between $0.27^{\prime\prime}$ and $1.8^{\prime\prime}.$ This ensures that known lenses are recovered, and chance alignments of quasars with other objects are safely rejected.
 By working directly with WISE-selected quasar-like objects, this problem is bypassed and isolated quasars are automatically rejected.

When applied to the Southern Galactic Hemisphere, with $b<-20$ and -70<dec<7, this gave $\approx6\times10^4$ objects, of which 31381 singlets and 770 multiplets in the DES-DR1 footprint.

\subsection{Gaia-DR2 Multiplets}
\label{multiplets}
The nominal resolution of Gaia has improved from $2^{\prime\prime}$ (DR1) to $0.4^{\prime\prime}$ (DR2), so a simple upgrade consists in searching again for Gaia-DR2 multiplets corresponding to WISE sources. 
%
There is indeed some improvement from DR1 to DR2: out of the $\approx6\times10^4$ Gaia-DR1 singlet `baroques' from above, a match with DR2 gave 2149 multiplets, of which 134 with more than two \texttt{source} entries.

However, some candidates and small-separation lenses from DR1 have disappeared in the passage to DR2 (e.g. the lens WGD0508, \citealt{Agnello17}). The same has been noticed in Spiniello et al. (2018, subm) on a smaller footprint (KiDS-DR3, 450~$\mathrm{deg}^2$). 
In general, the Gaia-specific deblending does not seem to have a well defined behaviour with separation and flux-ratios, as some small-separation lenses are recognized as multiplets while larger-separation lenses correspond to Gaia singlets (e.g. WGD0150, WGD0245; \citealt{Agnello17}). 

In order to make additional progress over the searches in Gaia-DR1, we also relaxed the preselection photometric criteria. From a WISE search, we selected extragalactic objects as
\begin{eqnarray}
W1-W2>0.2+\sqrt{(\delta W1)^{2}+(\delta W2)^{2}}\ ,\\
W1<17\ ,\ W3<12.4\ 
\end{eqnarray}
basically excluding stars ($W1-W2\approx0$) and fainter objects with unreliable catalog magnitudes. This gave $\mathcal{O}(10^6)$ objects with Galactic Latitude $b<-15.$ Of these, $\mathcal{O}(10^5)$ are recognized as multiplets in Gaia-DR2. However, many are clustered towards the Galactic disc and the Magellanic Clouds, which can be explained with the line-of-sight alignment of quasars with stars.
In order to exclude most quasar-star pair contaminants, we then impose a threshold on the ratio $\Sigma_{m}/\Sigma_{s}$ between multiplet and singlet densities. For the Gaia doubles, we retain only objects in regions where $\Sigma_{m}/\Sigma_{s}<0.2;$ for Gaia triplets and quadruplets, we relax it to $\Sigma_{m}/\Sigma_{s}<0.3.$ After the overdensity threshold cuts, $\approx17000$ multiplets remain, of which 8146 lie in the DES-DR1 public footprint.

\subsection{Spectroscopic Surveys}
\label{Spectrosc}

Some of the longest known quasar lenses, such as e.g. Q0957+561 \citep{Walsh79} and lenses from the Hamburg-ESO survey \citep{Wisotzki96}, were found in wide-field, prism spectroscopic surveys.
Over the last decade, searches in the SDSS and BOSS (\citealt{Inada12,More16}) relied on samples of objects that were classified as quasar based on fibre spectroscopy. Also, some lenses recently found in the DES (\citealt{Ostrovski17, Agnello17}) had pre-existing fibre spectra in the Anglo-Australian Observatory databases, where they had been targeted as quasars or galaxies and not recognized as lenses, due to low image resolution.

For this reason, we examined two surveys with moderate depth and uniform coverage over large areas in the Southern Hemisphere: the six-degree field galaxy survey (6dFGS, \citealt{Jones04}) and the 2QZ \citep{Boyle00}. 
These were also based on different preselection cuts, based on optical cuts in all-sky surveys of moderate depth, focusing on galaxy-like (6dFGS) or quasar-like (2QZ) objects. In this sense, then, they are also complementary to the WISE criteria that we have used in our photometric searches.

We restrict these samples to objects with pipeline redshifts $z_{s}>0.5,$ as expected from simulated samples (OM10) and our benchmark of CASTLES/SQLS known lenses. This gave 1625 objects from the 6dFGS and 7147 from 2QZ, over the DES footprint.

\subsection{X-Ray}
\label{xray}
Quasars can also be targeted in X-ray emission, due to lower dust attenuation in that spectral range. The recently vetted cross-match \citep{sal18}
the ROSAT all-sky survey \citep{bol16} and XMMslew survey \citep{geo11} with WISE then provides an alternative preselection of possible candidates.
Some lenses discovered in ROSAT are indeed known, such as RXJ0911.4+0551 \citep{bad97}, RXJ0921+4529 \citep{mun01}, and RXJ1131-1231 \citep{slu03}. A fainter lens, CY2201-3201 \citep{cas06} was found in the Chandra deep fields.

We then used the publicly available cross-match tables\footnote{http://www.mpe.mpg.de/XraySurveys/2RXS/} to pre-select X-ray objects. Also in this search, we excluded possible stars with the same cuts as above. This yielded 13547 cutouts from ROSAT-WISE, and 2294 from XMM-WISE.

\subsection{Radio}
\label{Radio}
Similarly to X-ray preselection, radio searches are less affected by reddening and by the presence of a foreground galaxy. As a consequence, they bypass the colour selection that can affect optical or IR searches. Indeed, quasar lens discoveries in the past have benefited from hemispheric surveys like the Parkes-MIT-NRAO survey \citep{gri93} and dedicated observations by the CLASS \citep{Myers03} and J-VLA \citep{kin99} searches.

Here, we use the Sydney University Molonglo Sky Survey \citep[SUMSS][]{mau03}, covering the Southern Hemisphere down to $\approx10$~mJy at $0.84$~GHz. We match the catalog with WISE, using a nearest-neighbour match with 10$^{\prime\prime}$ search radius, and retained extragalactic objects as above. This resulted in 38204 cutouts within the DES-DR1 footprint.

\section{Performance}
\label{performance}
The performance of each method can be quantified in different ways. One metric is given by completeness and purity with respect to previous searches. Another possibility is the number of candidates that are produced by each method over the same footprint. The final evaluation of performance is the purity of the candidate samples, once spectroscopic follow-up is performed. The candidates presented in this paper are still awaiting spectroscopic confirmation, and published samples of lenses and contaminants over the DES footprint are still too small to build meaningful statistics. Then for the scope of this paper, we discuss the recovery of known lenses over DES-DR1, and properties of candidate samples from different searches.

\begin{figure*}
\includegraphics[width=17cm]{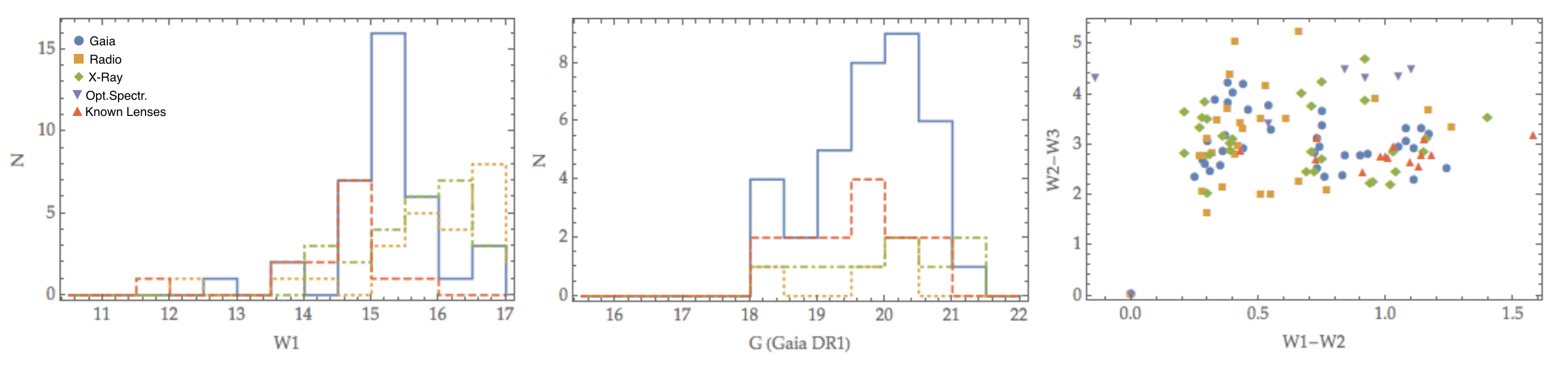}
 \caption{Magnitude and colour distributions of candidates and known lenses. \textit{Left:} $W1-$magnitude distribution for Gaia candidates (blue), radio candidates (orange, dotted), X-ray candidates (green, dot-dashed) and known lenses (red, dashed). \textit{Middle:} $G-$magnitude distribution, with the same colour coding. \textit{Right:} colours $W1-W2$ vs $W2-W3,$ same colour coding (and markers) as in the first panel.
BaROQueS and Multiplets are grouped, because they are both based on Gaia. Red triangles represent the known lenses recovered by the five searches.}
\label{fig:histogram}
\end{figure*}

\begin{table*}
\caption{Known lenses in the DES-DR1 public footprint recovered by the different search methods of Section~\ref{methods}. Magnitudes are from WISE and Gaia-DR1 and given in the Vega system. The first 6 systems in the table are the known quadruplets, whose cutouts are also shown in Figure~\ref{fig:quads}. Here, W2MG1 denotes 2MASS-vs-Gaia chromatic offsets (see Section 2.1), and `mult' or `sing' marks objects that correspond to Gaia multiplets or singlets, respectively.}
\label{tab:known_in_des}
\begin{center}
\begin{tabular}{l | c | c | c | c | c | l   }
\hline
 object & $W1$ & $W2$ & $W3$ & $G$ & methods & reference \\ 
\hline 
J00XX-20XX	&		14.56$\pm$0.03	&		14.01$\pm$0.04	&		10.76$\pm$0.10	&19.21	&		GaiaDR2mult	&	Lemon et al (in prep.) \\ 
J021416.37-210535.3	 & 	15.11$\pm$0.03	 & 	14.68$\pm$0.06	 & 	11.82$\pm$0.24	 & 	19.80	 & 	GaiaDR2mult	 & 	Agnello (2018, RNAAS) \\
J023233.16-211725.7	 & 	14.02$\pm$0.03	 & 	12.99$\pm$0.03	 & 	10.06$\pm$0.05	 & 	18.93	 & 	XMMslew	 & 	\citealt{Wisotzki99}	 \\
J025942.86-163543.0	 & 	13.86$\pm$0.03	 & 	13.13$\pm$0.03	 & 	10.03$\pm$0.05	 & 	19.22	 & 	W2MG1sing	 & 	\citealt{Schechter18}	 \\
J04XX-33XX	 & 	13.85$\pm$0.03	 & 	12.71$\pm$0.02	 & 	9.94$\pm$0.04	 & 	20.22	 & 	W2MG1mult,GaiaDR2mult	 & Anguita et al. (2018 subm.)	 \\
J040821.63-535358.9	 & 	14.08$\pm$0.02	 & 	13.17$\pm$0.02	 & 	10.75$\pm$0.06	 & 	19.63	 & 	GaiaDR2mult	 & 	\citealt{Lin17}	\\ J203342.12-472343.9	 & 	13.49$\pm$0.03	 & 	12.34$\pm$0.03	 & 	9.25$\pm$0.04	 & 	17.98	 & 	W2MG1mult,GaiaDR2mult	 & 	\citealt{Morgan04}	 \\ 
J203802.69-400813.9	 & 	11.43$\pm$0.02	 & 	10.25$\pm$0.02	 & 	7.49$\pm$0.02	 & 	19.44	 & 	W2MG1mult,GaiaDR2mult,6dFGS	 & 	Agnello et al 2017/8	 \\
\hline
J011557.37-524423.4	 & 	14.49$\pm$0.03	 & 	13.36$\pm$0.03	 & 	10.82$\pm$0.11	 & 	20.40	 & 	GaiaDR2mult	 & 	Agnello et al 2015	 \\
J014632.87-113339.2	 & 	14.17$\pm$0.03	 & 	13.08$\pm$0.03	 & 	10.45$\pm$0.06	 & 	18.39	 & 	W2MG1mult,GaiaDR2mult	 & 	Agnello et al 2017/8	 \\
J023527.42-243313.8	 & 	14.00$\pm$0.03	 & 	13.02$\pm$0.03	 & 	10.29$\pm$0.05	 & 	18.12	 & 	W2MG1mult,GaiaDR2mult	 & 	Agnello et al 2017/8	 \\
J051410.90-332622.4	 & 	13.29$\pm$0.03	 & 	11.71$\pm$0.02	 & 	8.54$\pm$0.02	 & 	17.80	 & 	HE, RASS	 & 	Gregg et al 2000	\\
J202139.38-411557.9	 & 	14.17$\pm$0.03	 & 	13.17$\pm$0.03	 & 	10.44$\pm$0.07	 & 	19.27	 & 	W2MG1mult,GaiaDR2mult	 & 	Agnello et al 2017/8	 \\
\hline
\end{tabular}
  \end{center}
\end{table*}

\subsection{Known Lenses, and the DES subsample}

Table~\ref{tab:known_in_des} lists the known lenses recovered over the DES-DR1 footprint. Five lenses are two image systems, and eight are (at least) quadruplets. Two of them (WG~04-33, WG~00-20) have been found independently within the STRIDES collaboration (Schechter \& Treu, private comm.) and are awaiting publication. For this reason, we list them as known lenses, with partially obscured coordinates, instead of as new candidates from our searches. The discovery of WG~0214-2105 has been recently reported by Agnello (2018, RNAAS).

The recovery statistics of known lenses in Gaia-DR2 is slightly better than for Gaia-DR1 (e.g. Agnello 2017). We quantify this through a list of 214 known quasar lenses and pairs collected from the SQLS \citep{Inada12} and CASTLES databases, consistently with previous work \citep{Williams17}. Of these, 43 are missing from Gaia-DR2. Of the 171 that are detected, 124 are resolved in two sources, 15 into three sources, and six into four sources. Small-separation lenses are absent in Gaia-DR2, because of a sharper cutoff at $\approx0.5^{\prime\prime}$ separations in the DR2 catalog. These include also lenses that were found thanks to Gaia-DR1 (e.g. WGD~0508, Agnello et al. 2017).

\subsection{Candidates}
Visual inspection of targets from the above methods resulted in: 17 BaROQuES in Gaia-DR1 vs 2MASS, of which 5 Gaia multiplets; 23 candidates from Gaia-DR2 multiplets; 7 candidates from fibre-spectroscopy (3 from 6dFGS and 4 from 2QZ); 30 candidates from X-ray preselection, of which 29 from ROSAT; and 29 candidates from SUMSS.
The full lists of high-graded candidates are provided in Tables~\ref{tab:cands1} and~\ref{tab:cands2}, divided by search method. We list their names (J2000 coordinates), as well as their WISE (W1,W2,W3) and Gaia-DR1 ($G$) magnitudes.

Figure~\ref{fig:histogram} shows the colour and magnitude distributions of the candidates. Different colours and markers represent the different methods through which the candidates have been identified. Known lenses are plotted in red (dashed in the magnitude histograms, triangles in the colour-colour plot). Gaia-selected candidates extend the regime of known lenses towards fainter magnitudes, still with a cutoff at the faint end due to noisy WISE magnitudes and the drop in Gaia completeness at $G\approx20.$ X-ray and radio selected candidates populate the faint end of the magnitude distribution, with few objects corresponding to optical detections in Gaia. 

Two main clusters of systems are visible in the WISE colour-colour plot: objects with high $W1-W2,$ compatible with QSO-dominated photometry, and systems with lower $W1-W2,$ compatible with galaxy-dominated photometry. Systems with low W1-W2 (left-most panel) mostly have $W3\geq$11.8, which is the threshold where WISE magnitudes become unreliable. There is a prevalence of radio-selected objects in the galaxy-dominated clump, whereas X-ray and Gaia selected systems are evenly distributed. Known lenses are mostly associated to the QSO-dominated regime.

\begin{figure*}
\includegraphics[width=18cm]{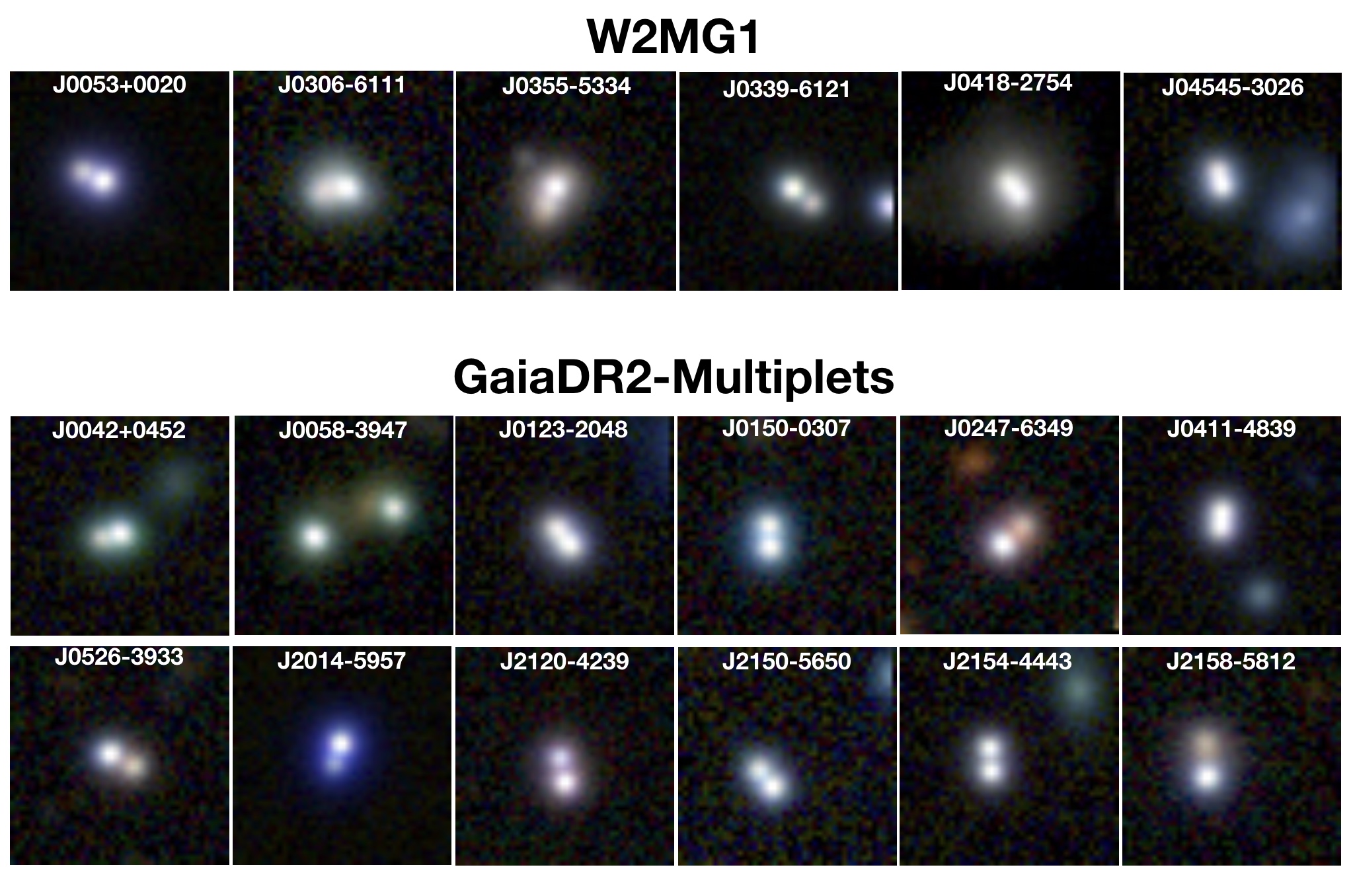}
 \caption{Examples of candidates found in Gaia DR1 and DR2, with WISE preselection. W2MG1 denotes WISE objects with acceptable offsets between their 2MASS and Gaia-DR1 catalog positions, with recomputed astrometry from our BaROQuES scripts.}
\label{fig:Candidates_multiplet}
\end{figure*}
\begin{figure*}
\includegraphics[width=18cm]{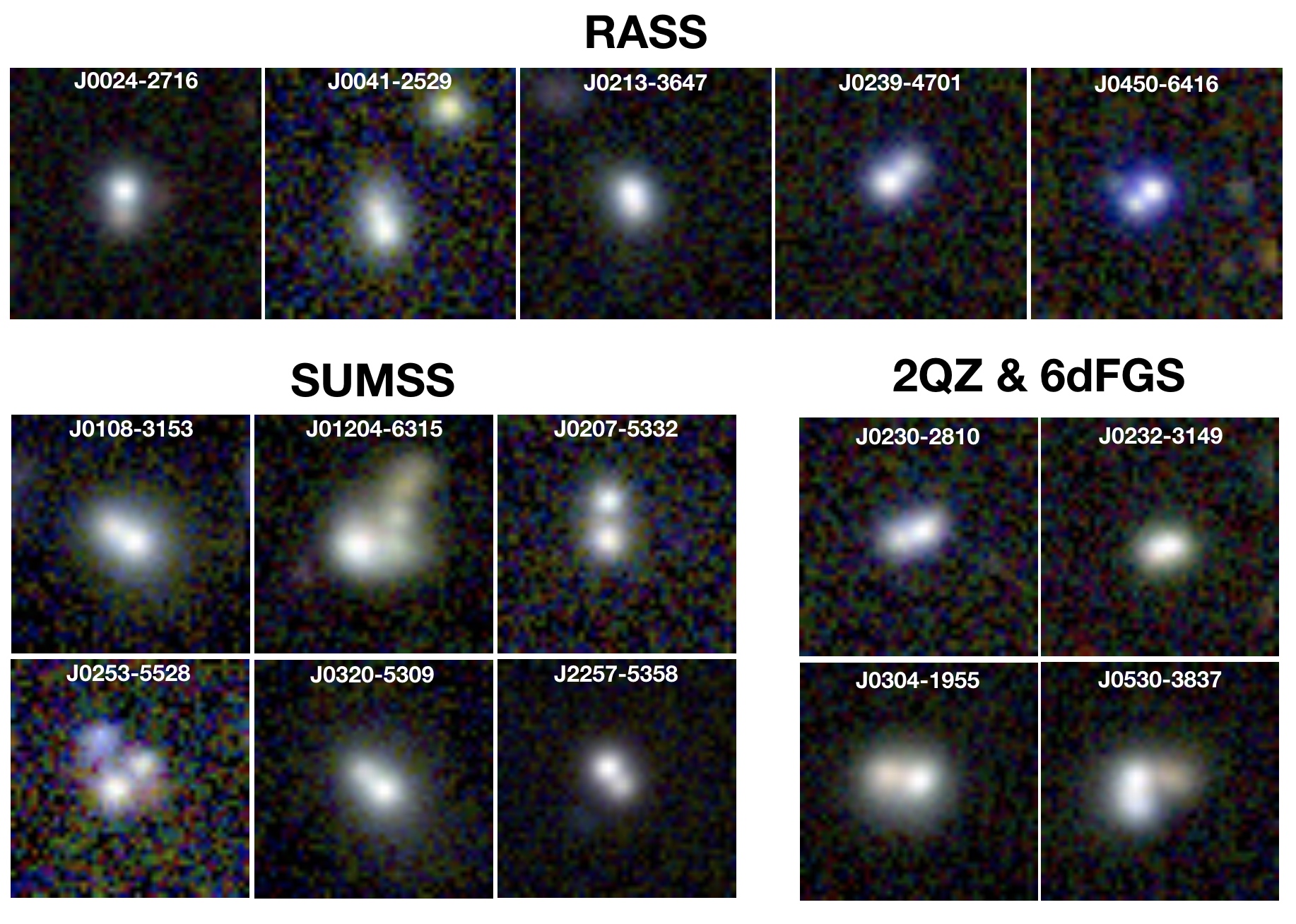}
 \caption{Examples of candidates found with spectroscopic (2QZ, 6dFGS), radio (SUMMS) or X-Ray (ROSAT, XMMslew) surveys.}
\label{fig:Candidates_others}
\end{figure*}

Candidates selected from each method are shown in Figures~\ref{fig:Candidates_multiplet} and~\ref{fig:Candidates_others}. All methods produce sensible results upon visual inspection, albeit with a predominance of high-grade candidates from the Gaia searches. Some candidates (mostly from SUMMS and spectroscopic surveys) may be lenses with bright deflectors, or mergers of compact narrow-line galaxies at lower redshift. With the depth ($i\approx22.5$) and image resolution of Pan-STARRS and DES-DR1 ($\approx0.26^{\prime\prime}/\mathrm{px},$ $\approx1^{\prime\prime}$ seeing), these contaminants are still present in candidate samples.

\subsection{Quadruplets}

Due to their higher number of images, with respect to doubles, quadruplets yield more constraints on the deflector potential and stellar mass fraction, and for this reason they are especially valued probes of cosmography \citep[e.g.][]{Refsdal64, Blandford92, Witt00, Suyu13, Treu16, Bonvin17} and microlensing studies \citep[e.g.][]{sch14}.

A quadruplet configuration requires an even closer alignment of source and deflector than doubles. The fraction of quadruplets in statistically-complete quasar lens samples is estimated at $14\%$ (OM10). 
The two quadruplets found within this search, WG210014.9-445206.4 and WG021416.37-210535.3, correspond to Gaia-DR2 triplets and are already shown in Figure~\ref{fig:quads}. 
Most of the other candidates, except some radio-selected and spectroscopically-selected candidates, are doubles. We emphasize that, while WG~0214-2105 requires little follow-up for confirmation, in principle the chromaticity in WG~2100-4452 requires a final step of spectroscopic confirmation to secure its lensing nature.

Two quadruplets found in this search, at J04-33 (Anguita et al., subm.) and J00-20 (Lemon et al., in prep.) have also been discovered independently within the STRIDES\footnote{STRIDES is a broad external collaboration of the Dark Energy Survey, \texttt{strides.astro.ucla.edu}} collaboration (Schechter \& Treu, private comm.). Even though they were found \textit{blindly} by our search, using exclusively public datasets, and so would be legitimate candidates in the framework of this paper, we prefer to list them as known lenses with partially obscured coordinates, in view of publication by other teams.

\subsection{Completeness}

The statistics of quadruplets is also an indication of completeness in a quasar lens search, relative to the number of objects at preselection. A `demanding' quasar preselection (e.g. $W1-W2>0.55+\sqrt{(\delta W1)^{2}+(\delta W2)^{2}},$ Agnello et al. 2017) produces $\approx40$ objects per square degree, which (adopting OM10 lensing rates) gives $\approx20$ lenses over the DES footprint, of which $\approx 2.8\pm1.67$ should be quads. A WISE preselection like the ones in this paper produces $110\pm10$ objects per square degree, and should result in $7.7\pm0.7$ quadruplets over the DES footprint. The current number of 9 quads (partly known and partly new) suggests that the steps of target and candidate selection are reasonably complete.

In order to find more quads, then, the main bottleneck is that of colour preselection. Searches with hybrid, optical-IR colours would allow one to relax the WISE extragalactic cuts, but in this exploration we preferred to use the DES-DR1 only for the final stage of visual inspection, in order not to impose a prejudice on the optical colours of lensed quasars upon our searches.

\section{Discussion}
\label{results}
We have discovered two previously unknown quadruplets, WG210014.9-445206.4 and WG021416.37-210535.3. We have independently discovered another two (J00-20, J04-33) that are currently awaiting publication by the STRIDES collaboration, which we prefer to list among other known lenses in the DES footprint.

The use of five different selection methods, over spectral ranges ranging from X-ray to radio through optical, has resulted in 100 quasar lens candidates over the DES-DR1 public footprint. The number of quadruplets found in this paper suggests that the searches are reasonably complete, and that the main bottleneck is the very first step of object preselection.

All of the searches involved a step of WISE pre-selection, imposing some loose cuts on colours in order to isolate extragalactic objects, and results in two main clusters of candidates based on whether their photometry is quasar-dominated or galaxy-dominated. Due to the current capabilities of the DESaccess server with large cutout queries, approximately two thirds of targets in the DES-DR1 footprint were properly displayed, so another $\approx50$ candidates may be still awaiting discovery.

\subsection{Deeper and Sharper WISE preselection}
At the stage of WISE preselection, we are currently limited by two factors: the uncertainties in $W1,W2,W3$ magnitudes; and image resolution. First, the searches are cut at $W1<17$ and $W2<15.6$ since the magnitudes of fainter objects are more affected by Earth-shine and become quickly unreliable. Second, the $\approx6^{\prime\prime}$ (FWHM) image resolution of WISE prevents any comparison of the mid-IR colours of different components, which at these wavelengths should be unaffected by reddening and microlensing and would then make a powerful selection criterion.

Forced photometry upon WISE cutouts, using object positions from sharper surveys (like Gaia or the DES), would alleviate these issues and enable a deeper and sharper search, maximizing the information content of WISE. This, of course, requires that multiple components are identified by the optical surveys, and that their separation is wide enough ($\gtrsim2.5^{\prime\prime}$) so that the magnitudes inferred through forced photometry are reliable. From a re-reduction \citep{mei17} and forward modeling of WISE/NeoWISE survey tiles, forced-photometry magnitudes are currently available for the SDSS \citep{lan16} and the DECaLS \citep{dey18} footprints, covering `Northern' declinations above $-20$~degrees. By using Gaia multiplets as a base for forced photometry, a search based on mid-IR (deblended) colours may be extended to the whole sky. However, the completeness of the resulting sample would be ill characterized, as the current releases of Gaia fail to properly deblend lenses (e.g. WGD~0150, WGD~0245, Agnello et al. 2017) that are clearly deblended into multiplets by other surveys such as the DES.

Another way of increasing depth and imaging resolution, in the BaROQuES approach, would be to replace 2MASS with deeper surveys, such as UKIDSS \citep{war07} or VHS \citep{mcm13}. However, their sky coverage is not homogeneous, whereas 2MASS covers the whole sky, and their application to a DES search is of limited use since their overlap with the DES footprint is only partial. Wile UKIDDS is primarily a Northern Hemisphere survey, VHS was planned to cover the whole Southern Hemisphere, but\footnote{Based on the final release (2018-02-09) by the ESO Archive Group, \texttt{https://www.eso.org/qi/catalog/show/217}, see `Description' for warnings on catalog usage and missing entries.} it is mostly limited to $-20\leq\mathrm{dec.}<0$ and $-50\leq\mathrm{dec.}<-40$, and not homogeneously.

\subsection{Towards LSST}
While near-IR surveys (e.g. VHS, UKIDSS, VHS) are ideally suited for chromatic offsets, due to the different colours of sources and deflectors, optical surveys can be a valid alternative. Lemon et al. (2018) have successfully used SDSS instead of 2MASS, and our BaROQuES procedures have been applied with success to the KiDS-DR3 450~deg$^2$ footprint (Spiniello et al. 2018, subm.). In the near future, LSST (SV due 2022, SciOps due 2023) will enable the extension of these searches to the whole Southern Hemisphere.

Despite the deluge of multi-epoch data from the full survey, the foreseen volume of nightly observations is still manageable, and automated procedures for chromatic offsets can be run on a nightly basis. This pipeline-oriented approach will then bring the discovery of lensed quasars to the same, high-cadence pace of transient discoveries. Ideally, chromatic offsets can be computed among different bands within LSST, without the recourse to external surveys like Gaia or 2MASS, provided nightly data are deep enough to guarantee robust astrometry in separate bands.

\subsection{Gaia DR3, DR4 and Euclid}
The DR2 of Gaia provided additional information over DR1, such as blue and red pass-band magnitudes ($G_{\rm{bp}},$ $G_{\rm{rp}}$), as well as parallax and proper motion, for sources towards the bright end. It has a safety cutoff at $\approx0.5^{\prime\prime}$ nearest neighbours, which prevents the discovery of small-separation lenses through multiplet recognition but may be reassessed in future releases. The pipeline proper motions of known lenses are not necessarily small (Spiniello et al. 2018, subm.), and colours from $G_{\rm{bp}},$ $G_{\rm{rp}}$ can only be used for the brighter candidates.

The final Gaia-DR4 release, due in 2020, is planned to provide low-resolution spectra of all detected Gaia sources. This would provide a selection of lens candidates based uniquely on Gaia, bypassing broad-band colour selection, through the detection of emission lines from the source quasars. This approach closely resembles the Hamburg-ESO strategy of isolating quasars based on their low-resolution spectra and following them up with higher imaging resolution, but it would be applied over the whole sky and down to fainter magnitudes.

Beyond the spectroscopic and multiplet selection in Gaia, an obvious step forward will be the Euclid space mission \citep{lau10}. The lessons learnt from the Gaia pipelines, in deblending objects into multiple sources and extracting spectra, will be essential to the development of lens searches within Euclid. In these all-sky perspectives, significant improvement is needed on the final stage of candidate selection, whether it is based on direct cutout modeling or direct visual inspection. Dedicated deblenders are being developed to this purpose \citep[e.g.][]{Chan15,Schechter18}, but they still require manual `tuning' to the specifics of each survey, and a considerable amount of visual inspection.

\subsection{Beyond ROSAT: eROSITA}
The optical and X-ray choices of preselection are quite complementary. On the one side, X-ray selected sources populate the faint end of the WISE and Gaia magnitude distribution, and sometimes miss Gaia \texttt{source} counterparts. On the other, only two of the known lenses in the DES are detected by our ROSAT-WISE search, and in general X-ray selected lenses are a minority among the CASTLES sample.

In the near future, eROSITA \citep{mer12} promises a $\approx20$ deeper coverage than ROSAT in the soft X-ray, and the first ever all-sky survey in the hard X-ray. The forecast of $\approx100$ detected objects per square degree, over the four-year span of the survey mission, is comparable to the WISE extragalactic selections used in this paper, and based on completely independent spectral signatures. In particular, from the expected number of active galactic nuclei (AGN) with redshift $1<z_{s}<3$ and the OM10 estimates, eROSITA should detect $\approx120$ lensed quasars over its 30000~deg$^2$ footprint, a quarter of which would have an obscured AGN as a source.

\section*{Acknowledgments} 
CS has received funding from the European Union's Horizon 2020 research and innovation programme under the Marie Sklodowska-Curie actions grant agreement n. 664931. 
CS wishes to thank N.~R.~Napolitano for interesting discussion about the lens searches. 
AA wishes to thank P.~L.~Schechter for support and encouragement, and gratefully acknowledges hospitality by the Harvard ITC, where the Gaia-DR2 searches were run.
\\
This research made use of the cross-match service provided by CDS, Strasbourg.
This research has made use of the NASA/IPAC Infrared Science Archive, which is operated by the Jet Propulsion Laboratory, California Institute of Technology, under contract with the National Aeronautics and Space Administration. 
This publication makes also use of data products from the Two Micron All Sky Survey, which is a joint project of the University of Massachusetts and the Infrared Processing and Analysis Center/California Institute of Technology, funded by the National Aeronautics and Space Administration and the National Science Foundation.\\
This project used public archival data from the Dark Energy Survey (DES).
 Funding for the DES Projects has been provided by the U.S. Department of Energy, the U.S. National Science Foundation, the Ministry of Science and Education of Spain, the Science and Technology Facilities Council of the United Kingdom, the Higher Education Funding Council for England, the National Center for Supercomputing Applications at the University of Illinois at Urbana-Champaign, the Kavli Institute of Cosmological Physics at the University of Chicago, the Center for Cosmology and Astro-Particle Physics at the Ohio State University, the Mitchell Institute for Fundamental Physics and Astronomy at Texas A{\&}M University, Financiadora de Estudos e Projetos, Fundacao Carlos Chagas Filho de Amparo a Pesquisa do Estado do Rio de Janeiro, Conselho Nacional de Desenvolvimento Cientifico e Tecnologico and the Ministerio da Ciencia, Tecnologia e Inovacao, the Deutsche Forschungsgemeinschaft and the Collaborating Institutions in the Dark Energy Survey.\footnote{See \texttt{https://des.ncsa.illinois.edu/thanks} for full list of Collaborating Institutions.}
%
Based in part on observations at Cerro Tololo Inter-American Observatory, National Optical Astronomy Observatory, which is operated by the Association of Universities for Research in Astronomy (AURA) under a cooperative agreement with the National Science Foundation.



\appendix
\section{List of high-graded candidates}
In the following tables we provide a list of high-graded lensed quasars candidates found with these searches over DES-DR1, divided by search method. The magnitudes are from WISE and Gaia-DR1 and given in the Vega system.

\begin{table*}
\caption{List of candidates recovered by the methods based on optical colours and spectra described in Section~\ref{methods}.}
\label{tab:cands1}
\begin{center}
\begin{tabular}{l | c | c | c | c | c | c | l }
\hline
 Object Name & $W1$ & $W2$ & $W3$ & $G$ & Methods  & Other \\
\hline
J000128.98-554959.1	&		14.93$\pm$0.03	&		14.17$\pm$0.04	&		11.85$\pm$0.23	&		19.39	&		W2MG1mult	&	 \\
J005301.98+002043.2	&		15.13$\pm$0.04	&		14.02$\pm$0.05	&		11.75$\pm$0.34	&		17.65	&		W2MG1sing	& PanSTARRS	 \\
J011455.07-054753.7	&		14.92$\pm$0.04	&		14.59$\pm$0.06	&		10.73$\pm$0.10	&		19.91	&		W2MG1mult	& PanSTARRS	 \\
J011509.25-231453.5	&		14.41$\pm$0.03	&		14.06$\pm$0.04	&		11.51$\pm$0.18	&		19.14	&		W2MG1sing	& PanSTARRS\\		
J020157.21-051000.9	&		14.79$\pm$0.03	&		14.35$\pm$0.04	&		11.47$\pm$0.15	&		19.92	&		W2MG1sing	& PanSTARRS\\	
J021524.18-472845.0	&		14.41$\pm$0.03	&		13.33$\pm$0.03	&		10.04$\pm$0.05	&		18.60	&		W2MG1sing	& \\	
J023207.66-325458.1	&		14.95$\pm$0.03	&		14.59$\pm$0.05	&		11.76$\pm$0.18	&		19.13	&		W2MG1mult	&	KiDS \\
J023639.92-475231.6	&		14.06$\pm$0.03	&		13.31$\pm$0.03	&		9.95$\pm$0.05	&		20.13	&		W2MG1sing	& \\	
J030606.05-611130.0	&		15.22$\pm$0.03	&		14.78$\pm$0.04	&		10.61$\pm$0.07	&		20.31	&		W2MG1sing	& \\
J033908.87-612144.7	&		14.61$\pm$0.03	&		13.44$\pm$0.03	&		10.26$\pm$0.05	&		18.45	&		W2MG1mult	&	 \\
J035542.64-533440.3	&		14.31$\pm$0.03	&		13.56$\pm$0.03	&		9.92$\pm$0.03	&		20.23	&		W2MG1sing	& \\
J040148.10-251438.0	&		14.17$\pm$0.03	&		13.12$\pm$0.03	&		10.20$\pm$0.06	&		18.79	&		W2MG1sing  & PanSTARRS\\
J041848.07-275410.1	&		14.34$\pm$0.03	&		13.88$\pm$0.03	&		10.21$\pm$0.05	&		19.85	&		W2MG1mult	& PanSTARRS	 \\
J045453.01-302636.0	&		14.85$\pm$0.03	&		14.48$\pm$0.04	&		11.33$\pm$0.13	&		18.79	&		W2MG1sing	& PanSTARRS \\
J050120.31-633247.8	&		13.47$\pm$0.02	&		12.54$\pm$0.02	&		9.75$\pm$0.03	&		20.58	&		W2MG1sing	& \\	
J215713.63-420149.5	&		12.40$\pm$0.02	&		11.16$\pm$0.02	&		8.65$\pm$0.02	&		17.84	&		W2MG1sing	& \\	
\hline
J003247.75-243429.0	&		16.39$\pm$0.06	&		15.99$\pm$0.15	&		11.98 &		19.89	&		GaiaDR2mult	&	 \\
J004254.16+045254.2	&		14.06$\pm$0.03	&		13.81$\pm$0.05	&		11.49 &					18.27	&		GaiaDR2mult	&	PanSTARRS \\
J005813.40-394724.6	&		14.77$\pm$0.03	&		14.05$\pm$0.04	&		11.23$\pm$0.12	&		19.42	&		GaiaDR2mult	&	PanSTARRS \\
J012325.25-204852.9	&		14.68$\pm$0.03	&		14.37$\pm$0.05	&		11.94 &					18.78	&		GaiaDR2mult	&	PanSTARRS \\
J015033.23-030746.5	&		15.97$\pm$0.05	&		14.83$\pm$0.07	&		11.53$\pm$0.20	&		20.07	&		GaiaDR2mult	&	 PanSTARRS\\
J024754.77-634923.2	&		15.10$\pm$0.03	&		14.36$\pm$0.04	&		11.44$\pm$0.11	&		19.68	&		GaiaDR2mult	&	 \\
J041137.60-483935.8	&		14.71$\pm$0.03	&		14.42$\pm$0.03	&		11.83$\pm$0.13	&		18.98	&		GaiaDR2mult	&	 \\
J052611.27-393346.6	&		15.47$\pm$0.04	&		14.74$\pm$0.05	&		11.65$\pm$0.16	&		19.98	&		GaiaDR2mult	&	 \\
J060832.98-351309.7	&		15.47$\pm$0.04	&		14.39$\pm$0.04	&		11.34$\pm$0.12	&		20.05	&		GaiaDR2mult	&	 \\
J212034.78-423904.7	&		14.82$\pm$0.03	&		13.71$\pm$0.04	&		10.83$\pm$0.11	&		19.37	&		GaiaDR2mult	&	 \\
J201425.41-595746.6	&		14.81$\pm$0.03	&		13.91$\pm$0.04	&		11.17$\pm$0.13	&		17.56	&		GaiaDR2mult	&	 \\
J202649.74-422818.6	&		14.83$\pm$0.04	&		13.99$\pm$0.05	&		11.24$\pm$0.16	&		19.70	&		GaiaDR2mult	&	 \\
J203348.67-593640.1	&		13.28$\pm$0.02	&		12.55$\pm$0.02	&		10.04$\pm$0.05	&		19.99	&		GaiaDR2mult	&	 \\
J210014.9-445206.4	&		14.14$\pm$0.03	&		13.42$\pm$0.03	&		10.74$\pm$0.10	&		18.79	&		GaiaDR2mult	&	 \\
J211242.15-595924.2	&		14.98$\pm$0.03	&		14.44$\pm$0.05	&		10.70$\pm$0.10	&		17.76	&		GaiaDR2mult	&	 \\
J212354.86+004416.2	&		16.04$\pm$0.06	&		15.66$\pm$0.14	&		11.87 &						-	&		GaiaDR2mult	&	PanSTARRS \\
J215021.33-565008.5	&		16.14$\pm$0.11	&		15.76$\pm$0.18	&		11.57  &				20.30	&		GaiaDR2mult	&	 \\
J215431.92-444302.0	&		14.63$\pm$0.03	&		14.35$\pm$0.05	&		11.68 &					19.18	&		GaiaDR2mult	&	 \\
J215837.29-581203.9	&		14.70$\pm$0.03	&		13.87$\pm$0.03	&		11.52$\pm$0.17	&		19.67	&		GaiaDR2mult	&	 \\
J220819.74-631500.9	&		15.00$\pm$0.03	&		14.70$\pm$0.06	&		11.66 &					19.31	&		GaiaDR2mult	&	 \\
\hline
J001947.71-293255.2	&		16.33$\pm$0.07	&		16.47$\pm$0.24	&		12.16 &					-	&		2QZ		&	KiDS/PanSTARRS		 \\
J003503.89-313203.2	&		17.73$\pm$0.20	&		16.68 &					12.36 &					-	&		2QZ		&	KiDS		 \\
J023023.47-281003.3	&		16.54$\pm$0.07	&		16.00$\pm$0.14	&		12.60$\pm$0.42	&		-	&		2QZ		&	KiDS/PanSTARRS		 \\
J023221.00-314913.1	&		-	-	&		-	-	&					-	-	&					-	&		2QZ	&			 \\
J025001.69-373244.7	&		14.58$\pm$0.03	&		13.74$\pm$0.03	&		9.26$\pm$0.03	&		-	&		6dFGS	&		 \\
J030403.03-195524.4	&		15.86$\pm$0.04	&		14.76$\pm$0.05	&		10.28$\pm$0.06	&		-	&		6dFGS	& 	PanSTARRS		 \\
J053009.04-383734.9	&		15.18$\pm$0.03	&		14.26$\pm$0.04	&		9.97$\pm$0.04	&		-	&		6dFGS	&		 \\
\hline

\end{tabular}
\end{center}
\end{table*}

\begin{table*}
\caption{List of candidates recovered by the methods described in Section~\ref{methods}, based on X-ray and Radio data.}
\label{tab:cands2}
\begin{center}
\begin{tabular}{l | c | c | c | c | c | c | l }
\hline
 Object Name & $W1$ & $W2$ & $W3$ & $G$ & Methods  & Other \\
\hline
J002453.30-271644.0	&		15.53$\pm$0.04	&		14.82$\pm$0.07	&		11.96$\pm$0.33	&		20.21	&	RASS	& KiDS/PanSTARRS	 \\
J004123.99-252944.9	&		15.16$\pm$0.04	&		14.76$\pm$0.06	&		11.66$\pm$0.20	&		-	&		RASS	& PanSTARRS		 \\
J004659.57+042039.0	&		15.53$\pm$0.05	&		15.23$\pm$0.12	&		11.72$\pm$0.42	&		-	&		RASS	& PanSTARRS		 \\
J005044.12-524856.5	&		14.53$\pm$0.03	&		13.78$\pm$0.03	&		11.06$\pm$0.13	&		20.75	&	RASS	&		 \\
J005911.01-015544.5	&		15.41$\pm$0.04	&		15.13$\pm$0.10	&		11.60$\pm$0.31	&		-	&		RASS	& PanSTARRS		 \\
J010515.03-625238.5	&		15.64$\pm$0.04	&		14.92$\pm$0.05	&		12.46$\pm$0.39	&		-	&		RASS	&		 \\
J011226.92-252032.2	&		14.55$\pm$0.03	&		13.86$\pm$0.04	&		11.41$\pm$0.15	&		-	&		RASS	& PanSTARRS		 \\
J014802.99-210841.6	&		16.07$\pm$0.06	&		15.80$\pm$0.15	&		12.47	&				-	&		RASS	& PanSTARRS		 \\
J020807.24-183910.9	&		14.42$\pm$0.03	&		14.12$\pm$0.04	&		12.08	&				-	&		RASS	& PanSTARRS		 \\
J021039.97-341941.8	&		15.34$\pm$0.04	&		15.03$\pm$0.07	&		12.23	&				19.41	&	RASS	& KiDS		 \\
J021307.50-364715.5	&		16.16$\pm$0.05	&		15.95$\pm$0.13	&		12.31	&				20.70	&	RASS	&		 \\
J023208.04-640931.6 &		16.13$\pm$0.05	&		15.92$\pm$0.10	&		13.11			&				&	RASS 	&		\\	
J023225.72-295737.1	&		17.43$\pm$0.15	&		16.03$\pm$0.14	&		12.51	&				-	&		RASS	& KiDS/PanSTARRS		 \\
J023919.80-470108.9	&		16.73$\pm$0.07	&		15.98$\pm$0.11	&		11.73$\pm$0.17	&		-	&		RASS	&		 \\
J030034.07-494237.4	&		13.76$\pm$0.02	&		12.73$\pm$0.02	&		9.89$\pm$0.05	&		18.09	&	RASS	&		 \\
J032354.51-104649.3	&		17.75$\pm$0.18	&		16.83			&		12.14	&				-	&		RASS	& PanSTARRS		 \\
J033514.21-600808.5	&		14.02$\pm$0.03	&		12.87$\pm$0.02	&		10.03$\pm$0.04	&		17.55	&	RASS	&		 \\
J033656.27-154756.5	&		15.55$\pm$0.04	&		15.16$\pm$0.07	&		12.14	&				-	&		RASS	& PanSTARRS		 \\
J034732.60-175324.0	&		15.03$\pm$0.03	&		14.67$\pm$0.05	&		11.52$\pm$0.20	&		-	&		RASS	& PanSTARRS		 \\
J035546.66-214818.5	&		15.56$\pm$0.04	&		14.85$\pm$0.06	&		11.08$\pm$0.12	&		-	&		RASS	& PanSTARRS		 \\
J043801.62-622942.9	&		15.85$\pm$0.04	&		15.46$\pm$0.07	&		12.58					-	& -	&	RASS	&		 \\
J044808.16-184642.3	&		15.36$\pm$0.04	&		14.44$\pm$0.05	&		10.58$\pm$0.07	&		-	&		RASS	& PanSTARRS	 \\
J045045.35-641642.8	&		16.74$\pm$0.05	&		15.70$\pm$0.07	&		13.24$\pm$0.46	&		-	&		RASS	&		 \\
J053248.87-391759.1	&		13.57$\pm$0.02	&		12.62$\pm$0.02	&		10.37$\pm$0.06	&		19.92	&	RASS	&		 \\
J215029.44-020016.9	&		15.18$\pm$0.04	&		14.89$\pm$0.08	&		11.05$\pm$0.15	&		-	&		RASS	& PanSTARRS		 \\
J225915.21-485056.2	&		15.50$\pm$0.04	&		14.83$\pm$0.06	&		10.81$\pm$0.11	&		-	&		RASS	&		 \\
J231409.88-421549.6	&		13.87$\pm$0.03	&		12.71$\pm$0.03	&		9.60$\pm$0.04	&		-	&		RASS	&		 \\
J235111.37-375323.5	&		14.79$\pm$0.03	&		13.77$\pm$0.04	&		11.57$\pm$0.19	&		18.86	&	RASS	&		 \\
J055640.64-611521.3	&		14.76$\pm$0.03	&		13.82$\pm$0.03	&		11.58$\pm$0.10	&		19.50	&	XMMslew	&		 \\
\hline
J000806.23-561551.9	&		16.26$\pm$0.06	&		15.87$\pm$0.13	&		11.50$\pm$0.17	&		-	&		SUMSS	&		 \\
J010844.90-315335.7	&		14.87$\pm$0.03	&		14.59$\pm$0.05	&		12.56	&				-	&		SUMSS	& KiDS		 \\
J012043.82-631556.5	&		13.95$\pm$0.03	&		12.69$\pm$0.03	&		9.36$\pm$0.03	&		-	&		SUMSS	&		 \\
J013506.97-412612.0	&		11.91$\pm$0.03	&		11.25$\pm$0.02	&		6.04$\pm$0.01	&		17.87	&	SUMSS	&		 \\
J020722.82-533224.1	&		15.72$\pm$0.04	&		15.21$\pm$0.06	&		11.71$\pm$0.17	&		-	&		SUMSS	&		 \\
J022238.34-531958.2	&		16.07$\pm$0.05	&		15.64$\pm$0.09	&		12.23 	&				-	&		SUMSS	&		 \\
J025310.73-552823.2	&		16.13$\pm$0.05	&		15.75$\pm$0.09	&		12.07$\pm$0.23	&		-	&		SUMSS	&		 \\
J032028.96-530944.7	&		14.80$\pm$0.03	&		14.50$\pm$0.04	&		12.87	&				-	&		SUMSS	&		 \\
J035425.21-642344.5	&		13.12$\pm$0.02	&		11.95$\pm$0.02	&		8.30$\pm$0.02	&		-	&		SUMSS	&		 \\
J042936.07-582237.5	&		16.05$\pm$0.03	&		15.71$\pm$0.05	&		12.24$\pm$0.15	&		-	&		SUMSS	&		 \\
J043228.41-545853.8	&		15.59$\pm$0.03	&		14.82$\pm$0.04	&		12.76$\pm$0.24	&		-	&		SUMSS	&		 \\
J044242.88-481249.5	&		16.12$\pm$0.05	&		15.70$\pm$0.09	&		12.75	&				-	&		SUMSS	&		 \\
J045947.39-641538.9	&		15.39$\pm$0.03	&		14.84$\pm$0.04	&		12.84$\pm$0.35	&		19.76	&	SUMSS	&		 \\
J050309.96-311652.0	&		15.89$\pm$0.04	&		15.62$\pm$0.08	&		12.88	&				-	&		SUMSS	&		 \\
J050709.35-465931.8	&		16.18$\pm$0.05	&		15.77$\pm$0.09	&		12.98	&				-	&		SUMSS	&		 \\
J051052.38-520129.2	&		15.34$\pm$0.03	&		14.98$\pm$0.05	&		12.84	&				19.31	&	SUMSS	&		 \\
J053128.21-383954.3	&		17.91$\pm$0.18	&		17.50$\pm$0.44	&		-	&					-	&		SUMSS	&	 \\
J211006.51-484121.1	&		16.21$\pm$0.06	&		15.77$\pm$0.12	&		12.49$\pm$0.45	&		-	&		SUMSS	&		 \\
J215533.59-483645.2	&		16.74$\pm$0.09	&		16.21$\pm$0.19	&		12.06$\pm$0.35	&		-	&		SUMSS	&		 \\
J225710.84-535824.9	&		15.45$\pm$0.04	&		15.13$\pm$0.08	&		12.32$\pm$0.42	&		19.59	&	SUMSS	&		 \\
J222136.92-574847.6	&		15.10$\pm$0.04	&		14.44$\pm$0.04	&		12.20$\pm$0.33	&		-	&		SUMSS	&		 \\
J233432.25-585646.7	&		15.71$\pm$0.04	&		15.41$\pm$0.09	&		12.32$\pm$0.42	&		-	&		SUMSS	&		 \\
J234038.36-461254.8	&		16.27$\pm$0.06	&		15.66$\pm$0.11	&		12.17$\pm$0.33	&		-	&		SUMSS	&		 \\
J234721.48-561943.7	&		15.46$\pm$0.04	&		15.17$\pm$0.06	&		12.42	&				-	&		SUMSS	&		 \\
J234839.67-520936.5	&		14.61$\pm$0.03	&		14.10$\pm$0.04	&		12.10$\pm$0.34	&		-	&		SUMSS	&		 \\
J235311.68-544823.3	&		16.80$\pm$0.09	&		15.84$\pm$0.13	&		11.96	&				-	&		SUMSS	&		 \\
\hline
\end{tabular}
\end{center}
\end{table*}


\bsp	
\label{lastpage}
\end{document}